\documentstyle[prb,aps,multicol,epsf]{revtex}
\begin{document}
\draft
%%%%%%%%%%%%%%%%%%%%%%%%%%%%%%%%%%%%%%%%%%%%%%%%%%%%%%%%%%%%%%%%%%%%%%%%%%
\title{Microscopic model, spin wave theory, and competing orders in   
double perovskites}
 \author{G. Jackeli\cite{byline}} 
 \address{ Institut Laue Langevin, B. P. 156, F-38042,
 Grenoble, France} 
\maketitle
%%%%%%%%%%%%%%%%%%%%%%%%%%%%%%%%%%%%%%%%%%%%%%%%%%%%%%%%%%%%%%%%%%%%%%%%%
\begin{abstract}
%\widetext 
We present a microscopic theory of carrier-induced ferrimagnetism in metallic 
double perovskite compounds such as  
${\rm Sr}_{2}{\rm FeMoO}_{6}$ and ${\rm Sr}_{2}{\rm FeReO}_{6}$ 
which have recently attracted intense interest for their possible applications 
to magnetotransport devices. 
The theory is based on an effective "Kondo-like" Hamiltonian treated here 
within the large-$S$ expansion.
We find that depending on the value of the carrier density
the  ground state is  either a ferrimagnet or a layered antiferromagnet.
The ferrimagnetic state has a robust half-metallic electronic structure.
The transition to antiferromagnetic phase is first order accompanied
with the regime 
of phase separation.
We study  spin wave spectrum including quantum corrections
and find strongly enhanced quantum effects in the vicinity of zero-temperature
phase transition.

\end{abstract} 
%%%%%%%%%%%%%%%%%%%%%%%%%%%%%%%%%%%%%%%%%%%%%%%%%%%%%%%%%%%%%%%%% 
\begin{multicols}{2} 
\narrowtext 

The recent discovery of the room temperature magnetoresistance in 
double perovskite compounds \cite{cmr} has generated intense interest in
these materials  because of their potential importance as 
magnetotransport devices  as well as their rich and 
challenging properties.  
For applications   metallic ferrimagnetic (FiM)
compounds such as
${\rm Sr}_{2}{\rm FeMoO}_{6}$ and ${\rm Sr}_{2}{\rm FeReO}_{6}$ are of
particular interest, because both have  high magnetic transition
temperature ($T_{\rm c}\simeq 420$, and $\simeq 400$ K, respectively) and
their electronic structures have been suggested to be half-metallic.\cite{cmr}
Despite an enormous interest in half-metallic ferromagnets,
most of the  theoretical studies until now have been
only by   {\it ab initio} methods.\cite{sar,fan,sol,sai} 

In this letter we formulate a microscopic theory of the carrier-induced
ferrimagnetism in this family of magnetic compounds. 
The theory is based on a minimal model Hamiltonian, in which only the lowest
energy charge fluctuations, coupled to the local moments, 
are retained, and those which are moved to high
energies due to  strong correlations, are disallowed.
The model has two parameters and allows for reliable analytical
treatment of the physically relevant situation.

In the double perovskite structure Fe and Mo/Re  ions form  two face centered
cubic  sublattices and each pair of nearest-neighbor  
lattice sites is occupied by two different ions. 
In the ionic picture,  ${\rm Fe}$ is in the $3+$ valence
state and  half-filled $d$-shell forms a $S=5/2$ local moment. 
The Mo  is in $5+$ state
with one $t_{2g}$ electron
in $4d$-shell.\cite{sar} 
This extra electron, which  hybridizes with the same orbital states at
the neighboring  iron  sites, is responsible for  the system's lowest energy charge
excitations.\cite{charge} 
In the classical picture, only a spin down electron,  
with respect to a neighboring  
Fe local moment,
can hop to this Fe site. 
For 
a spin up electron this hopping is blocked by the Pauli principle, as 
 at the the iron  site  the spin up states are already all occupied.
Thus, {\it it is the exclusion principle}, which couples 
the itinerant and the local spins {\it kinematically}. 
This contrasts with the case of double-exchange  systems, in which
the intra-atomic Hund's rule is responsible for the coupling between 
the two subsystems.\cite{S} 
In the magnetic double perovskites, as in double exchange magnets, 
ordering of local moments occurs due to the reduction of the 
kinetic energy.
The kinetic energy in the former case is minimized
when the local moments are parallel to each other and antiparallel to
itinerant spins, and this causes   the ferrimagnetism.

{\it Model Hamiltonian.} --- 
We now turn to a  minimal model  Hamiltonian  reflecting 
the above discussion. We retain only low energy charge excitations in the system,
which  correspond to the  processes 
$({\rm Fe}~d^{5},{\rm B}^\prime~d^{n})\leftrightarrow({\rm
  Fe}~d^{6},{\rm B}^\prime~d^{n-1})$ 
(${\rm B}^\prime$ stands   
for {\rm Mo}/Re). 
The ${\rm Fe}$ ions are treated as  
localized spins $S=5/2$ and extra electrons from nonmagnetic Mo/Re ions
as charge carriers. There are $n=1$ and $n=2$ carriers per unit cell
in Mo and Re compounds, respectively.  
When a carrier is placed at a  site with core spin $S=5/2$,
the  total spin ${\cal S}$ can 
take two possible values ${\cal S}=2$ and ${\cal S}=3$.
However, the maximal allowed spin for 
six electrons in a $d$-shell is ${\cal S}=2$.
To project out the ${\cal S}=3$ spin state  
 we introduce  infinite local antiferromagnetic (AFM) coupling 
$J\rightarrow \infty$
between  
core  and itinerant spins.\cite{nagaev}
The carriers  occupy three degenerate $t_{2g}$  
($d_{xy},d_{yz},d_{yz}$) orbital states. 
In the cubic lattice the $t_{2g}$-transfer matrix is diagonal
in the orbital space and is non-zero only in the corresponding plane.
A minimal Hamiltonian can be written as a sum of three two
dimensional (2D) terms
${\cal H}=H_{xy}+H_{xz}+H_{yz}$, where each term corresponds 
to a given orbital state. All three terms have the same form given by
(we further omit orbital index for simplicity)
%%%%%%%%%%%%%%%%%%%%%%%%%%%%%%%%%%%%%%%%%%%%%%%%%%%%%%%%%%%%%%%
\begin{eqnarray}
H&=&-t\!\sum_{\langle ij\rangle\sigma}\left[
d_{i\sigma }^{\dagger}{{\bar d}}_{j\sigma}+{\rm H.c.}\right]
+
\sum_{i}\Delta_{\rm CT} n_{i}-\sum_{j}\Delta_{\rm CT} {\bar n}_{j}\nonumber\\ 
&+&\sum_{i}J\left[
{\bf S}_i{\bf s}_{i}-
An_i\right],
\label{H1}
\end{eqnarray}
%%%%%%%%%%%%%%%%%%%%%%%%%%%%%%%%%%%%%%%%%%%%%%%%%%%%%%%%%%%%%%%%%%
where the first term describes an electron hopping between 
 nearest-neighbor  Fe and Mo/Re ions, labeled by $i$ and $j$, respectively.
The operators $d_{i\sigma}$ ($n_i$) and   ${\bar d}_{j\sigma}$ (${\bar n}_j$)
corrsepond to  Fe and Mo/Re sublattices, respectively,
$2\Delta_{\rm CT}=E({\rm Fe}d^6,B^\prime d^{0})-E({\rm Fe}d^5,B^\prime d^{1})$
is a charge transfer gap, and 
${\bf S}_i$ (${\bf s}_{i}$) stands for the localized (itinerant) spin.
The last term with   $A=(S+1)/2$
 is chosen so that the exchange energy
vanishes for a state with ${\cal S}=2$.

We now consider  the ferromagnetic (FM)
 arrangement of local moments, which corresponds 
to the ground state of Hamiltonian (\ref{H1}) 
for carrier density $n<n_{1}$ (see below).
In the limit  $J\rightarrow \infty$ , we restrict the Hilbert space to spin
eigenstates with maximum allowed spin ${\cal S}$ at a given site.
Thus, the magnetic ion is in a state either with ${\cal S}=S$, 
without an extra electron, or of ${\cal S}=S-1/2$, 
with an extra electron.
The site states are labeled by $|{\cal S},M\rangle$ and 
$M$ is $z$-projection 
of the spin ${\cal S}$. Considering low lying excitations we can retain  
only the states
with $M={\cal S}$ and $M={\cal S}-1$. 
Moerover, as the value of the core spin is relatively large $S=5/2$ 
we perform the $1/S$  expansion around the ordered state.
Then the local Hilbert space consists
of four states:$|S,S\rangle$, $|S,S-1\rangle$, 
 $|S-\frac{1}{2},S-\frac{1}{2}\rangle$, and  
$|S-\frac{1}{2},S-\frac{3}{2}\rangle$.
To describe  transitions between these
states we introduce magnon
and fermion operators as  $B^{\dagger}|{\cal S},M\rangle=|{\cal S},M-1\rangle$ and
$D^{\dagger}_{\downarrow}|S,M\rangle=|S-\frac{1}{2},M-\frac{1}{2}\rangle$, respectively. 
Through corresponding matrix elements, one can find the following  relations  
$d_{\downarrow}=D_{\downarrow}\left[1-B^\dagger B/4S-B^\dagger
B/(32S^2)\right]$,  and  $d_{\uparrow}=-D_{\downarrow}B^\dagger/\sqrt{2S}$ 
to order $1/S^2$.
The state generated by  $D_{\uparrow}$ corresponds  to 
 ${\cal S}=S+1/2$ and  is projected out.
 
As a next step, we express the Hamiltonian (\ref{H1}) in
terms of the new operators, diagonalize its  band part, and arrive 
in the momentum
space to the following expression $H=H_{0}+H_{1}+H_{2}$, where
%%%%%%%%%%%%%%%%%%%%%%%%%%%%%%%%%%%%%%%%%%%%%%%%%%%%%%%%%%%%%%%%%%%%
%\end{multicols}
%\begin{widetext}
\begin{eqnarray}
H_{0}&=&\sum\limits_{{\bf k}}\left\{
E_{{\bf k}}[a^{\dagger}_{{\bf k}\downarrow}a_{{\bf k}\downarrow}
-b^{\dagger}_{{\bf k}\downarrow}b_{{\bf k}\downarrow}]
-\Delta_{\rm CT} {\bar d}^{\dagger}_{{\bf k}\uparrow}{\bar d}_{{\bf k}\uparrow}\right\},\nonumber\\
H_{1}&=&\frac{1}{\sqrt{2SN}}\sum_{{\bf k,\bf q}}\left
\{{\bar d}^{\dagger}_{{\bf k-q}\uparrow}[N_{{\bf k},{\bf q}}
a_{{\bf k}\downarrow}
+M_{{\bf k},{\bf q}}
b_{{\bf k}\downarrow}]
B_{\bf
q}^{\dagger}+\text{H.c.}\right\},\nonumber\\
H_{2}&=&\frac{1}{4SN}\left[1+\frac{1}{8S}\right]\sum_{{\bf k,\bf p,\bf q}}B_{\bf
q}^{\dagger}B_{{\bf q}+{\bf k}-{\bf p}}\left\{P^{aa}_{{\bf k},{\bf
    p}}a^{\dagger}_{{\bf k}\downarrow}a_{{\bf p}\downarrow}\right.\nonumber\\
&+&\left.P^{bb}_{{\bf k},{\bf
    p}}b^{\dagger}_{{\bf k}\downarrow}b_{{\bf
    p}\downarrow}
+P^{ab}_{{\bf k},{\bf
    p}}a^{\dagger}_{{\bf k}\downarrow}b_{{\bf
    p}\downarrow}+P^{ab}_{{\bf p},{\bf
    k}}b^{\dagger}_{{\bf k}\downarrow}a_{{\bf
    p}\downarrow}\right\},
\label{H2}
\end{eqnarray}
%\end{widetext}
%\begin{multicols}{2}
%%%%%%%%%%%%%%%%%%%%%%%%%%%%%%%%%%%%%%%%%%%%%%%%%%%%%%%%%%%%%%%%%%%%
where $a_{{\bf k}\downarrow}=u_{\bf k}D_{{\bf k}\downarrow}-
v_{\bf k}e^{ik_{x}}{\bar d}_{{\bf k}\downarrow}$,
 $b_{{\bf k}\downarrow}=v_{\bf k}D_{{\bf k}\downarrow}+
u_{\bf k}e^{ik_{x}}{\bar d}_{{\bf k}\downarrow}$, and 
$u(v)_{\bf k}=\sqrt{[1\pm \Delta_{\rm CT}/E_{\bf k}]/2}$.
The first term $H_{0}$, corresponding to the classical
 limit $(S\rightarrow\infty)$, 
describes  the band structure of the system. 
The electronic structure of each
 orbital state  is
 composed of the three bands: bonding ($b_{{\bf k}\downarrow}$)  and
 antibonding ($a_{{\bf k}\downarrow}$)  of down
spin electrons,  
and nonbonding (${\bar d}_{{\bf k}\uparrow}$) of Mo/Re up spin electrons.
 $E_{\bf k}=\sqrt{\varepsilon_{\bf k}^{2}+\Delta_{\rm CT}^{2}}$ and 
$\varepsilon_{\bf k}=2t(\cos k_x+\cos k_y)$ for the $d_{xy}$-orbital.
For $n$ carriers  per unit cell each bonding band of a given 
 orbital is $n/3$-filled and fully polarized. 
Given this band structure, the system
is half-metallic for arbitrary model parameters ($t$, $\Delta_{\rm CT}$) and 
for any realistic value of carrier density.
The last two terms $H_{1}$ and $H_{2}$ [see Eq.(\ref{H2})] 
are due to  the quantum nature of the core
spins. Spin up electrons, localized in the classical picture,
can hop thanks to the quantum nature of the local moments, but leaving
a trace of spin deviations along their trajectories.   
This process is  described by $H_{1}$. The last term 
$H_{2}$  describes fluctuations of the local moment
generated by the hopping of spin down electrons.
The corresponding vertices are given by
$N_{\bf k, q}=\varepsilon_{\bf k-q}u_{\bf k}$, 
$M_{\bf k, q}=\varepsilon_{\bf k-q}v_{\bf k}$, 
$P^{aa}_{{\bf k},{\bf p}}=-\varepsilon_{\bf k}v_{\bf k}u_{\bf
  p}-\varepsilon_{\bf p}u_{\bf k}v_{\bf p}$, $P^{bb}_{{\bf k},{\bf
    p}}=\varepsilon_{\bf k}u_{\bf k}v_{\bf p}+\varepsilon_{\bf
  p}v_{\bf k}u_{\bf p}$, and $P^{ab}_{{\bf k},{\bf p}}=\varepsilon_{\bf p}u_{\bf p}u_{\bf k}-\varepsilon_{\bf k}v_{\bf k}v_{\bf p}$.

{\it Spin wave theory.} -- We now turn to analysis of the spin wave spectrum. 
The magnon Green function  is  obtained from the Dyson equation
$D_{{\bf q},\omega}=[\omega-\Sigma_{{\bf q},\omega}]^{-1}$. 
Where magnon self-energy $\Sigma=\sum_{i}\Sigma_{i}$ and $i=xy,xz,yz$ refers
to a given orbital channel. We note that to the order we consider
the magnon self-energy does not mixes the different orbital channels.
This is  due to the fact that the electron propagator as well as 
the magnon-electron vertecies are diagonal in the orbital space.
To order $1/S$, the self-energy  $\Sigma$ includes  the second order perturbative
contribution from $H_{1}$ as well as first  order contribution from  $H_2$ 
(see first two diagrams in Fig 1.)
(The analytical form of $\Sigma$ is  lengthy and
will be presented elsewhere.) Here we discuss the results.
There are two branches of spin excitations, like those of localized 
ferrimagnets,
the gapless Goldstone mode 
and  the optical mode.\cite{ferri} In addition to this,
there  exists a Stoner continuum of spin-flip particle-hole excitations.
%%%%%%%%%%%%%%%%%%%%%%%%%%%%%%%%%%%%%%%%%%%%%%%%%%%%%%%%%%%%%%%%%%%%%
\begin{figure}
\epsfysize=40mm
\centerline{\epsffile{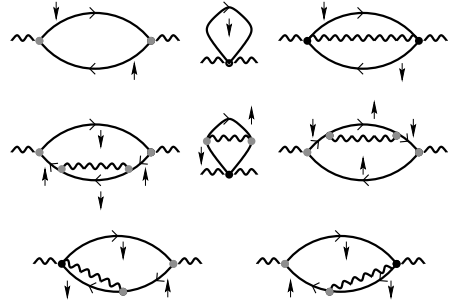}}
\caption{The leading order ($1/S$) and next-to-leading order ($1/S^2$)   
spin wave self-energies. Wave (solid) line stands for the
magnon (electron) propagator. The small arrow denotes an
electron spin and  gray (black) dot stands for the magnon-electron vertex
proportional to $1/\sqrt{S}$ ($1/S$).}
\label{f1}
\end{figure}
%%%%%%%%%%%%%%%%%%%%%%%%%%%%%%%%%%%%%%%%%%%%%%%%%%%%%%%%%%%%%%%%%%%%%
We below focus on the low energy mode.
The spectrum of the magnons at the quasi-classical level is Heisenberg-like 
$\omega_{\bf q}=16SJ_{1}[1-\gamma_{1\bf q}]+16SJ_{2}[1-\gamma_{2\bf q}]$,
where  $J_{1}$ and  $J_{2}$ are nearest  and next-nearest 
neighbor exchange couplings, respectively,
$\gamma_{1\bf q}=(\cos q_{x}\cos q_{y}+\cos q_{x}\cos q_{z}+\cos
q_{y}\cos q_{z})/3$, and $\gamma_{2\bf
  q}=(\cos^{2}\!q_{x}+\cos^{2}\!q_{y}+\cos^{2}\!q_{z})/3$. 
The carrier-induced exchange energies are
given by $J_{1(2)}=t^2/(16S^2N)\sum_{{\bf k}}{\bar \gamma}_{1(2)}n_{\bf k}/E_{\bf
  k}$, where ${\bar \gamma}_{1}=2\cos k_{x}\cos k_{y}$,
${\bar \gamma}_{2}=\cos 2k_{x}+\cos 2k_{y}$ and 
$n_{\bf k}$ is a Fermi distribution function.
We also note that there is no Landau damping of spin waves at this order.

\begin{figure}
\epsfysize=45mm
%\hspace*{0.1cm}
\centerline{\epsffile{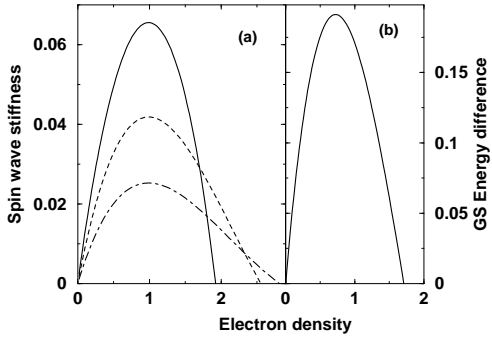}}
\caption{(a) Spin wave stiffness (in units of $t=1$)
versus carrier density, for different values of $\Delta_{\rm CT}$ ($\Delta_{\rm CT}=$0, 4, and
8, solid, dashed and dotted-dashed
lines, respectively); (b) Ground state (GS) 
energy difference  (in units of $t$, and for $\Delta_{\rm CT}=0$) between AFM-II
and ferrimagnetic states versus carrier density.}
\label{f2}
\end{figure}

In Fig. \ref{f2}(a) we show  carrier density  dependence of the   
spin stiffness ${\cal D}=16S(J_1+J_2)$ 
for  different values of $\Delta_{\rm CT}$.
At a fixed density the stiffness scales with the bandwidth:
${\cal D}\sim t$ for $t\gg\Delta_{\rm CT}$
and ${\cal D}\sim t^2/\Delta_{\rm CT}$ for $t\ll\Delta_{\rm CT}$.
As a function of density the spin stiffness exhibits strongly nonmonotonic
behavior. It 
shows a  maximum at optimal filling $n_{opt}\simeq 1$ and
then drops to zero at some critical density  $n_{cr}$.\cite{millis}

We now analyze quantum corrections to the harmonic spectrum discussed above.
These corrections are generated by the magnon self-energy diagrams
proportional to $1/S^2$.\cite{gol} 
They are governed by fermionic excitations 
and are thus different from those in insulating magnets.
At this order magnon self-energy includes fourth and second 
order contributions from
$H_{1}$ and $H_2$, respectively, as well as mixed type terms.
Graphically  they are presented in Fig. \ref{f1}. 
 From the structure of the diagrams one easily verifies that 
renormalization of the electron propagator due 
to the scattering of spin waves as well as 
all relevant vertex corrections are consistently taken into account at this
order. After collecting all the diagrams in this order one recovers the
Goldstone theorem, that illustrates the 
 consistency of the approximation.
%%%%%%%%%%%%%%%%%%%%%%%%%%%%%%%%%%%%%%%%%%%%%%%%%%%%%%%%%%%%%%%%%%%%%
%%%%%%%%%%%%%%%%%%%%%%%%%%%%%%%%%%%%%%%%%%%%%%%%%%%%%%%%%%%%%%%%%%%%%

The quantum corrections generate  
 two main effects in the magnon spectrum:
(i) they give rise to spin wave damping in the ground state,
and (ii) they modify 
the  semi-classical dispersion law of magnons.
The first nonzero contribution to magnon damping $\Gamma_{\bf q}$ is
proportional to $1/S^3$ and is due to the spin wave scattering of 
carrier density fluctuations. To evaluate damping, one has to replace 
the bare magnon propagator in the self-energy diagrams  
by that obtained in  harmonic approximation 
(see Golosov in Ref.\onlinecite{gol}). In the long wave-length limit
one finds $\Gamma_{\bf q}\propto  \langle f_{k_{\rm F}}\rangle tq^6/S^3$,
where $\langle f_{k_{\rm F}}\rangle$ denotes the Fermi surface average
of $f_{\bf k}=(\sin{k_x}+\sin{k_y})^2$.

\begin{figure}
\epsfysize=50mm
\centerline{\epsffile{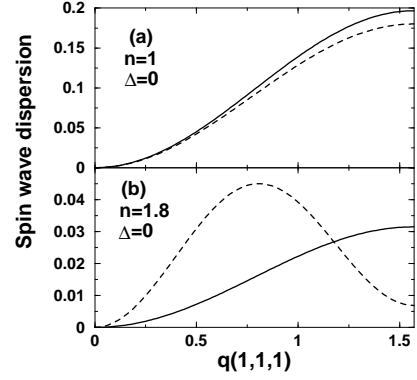}}
\caption{Spin wave spectrum  in $[111]$ direction 
(in units of $t=1$) versus momentum.  
Solid and dashed lines represent the harmonic and renormalized 
 magnon dispersion law, respectively.} 
\label{f3}
\end{figure}

Figure  \ref{f3} shows magnon dispersion obtained semiclassically (solid line) 
and with quantum
 corrections included (dashed line),  for carrier densities 
$n=1$, Fig.\ref{f3}(a), and $n=1.8$, Fig.\ref{f3}(b).
In the case of optimal carrier density, $n=1$, 
the quantum effects are small and practically do not alter harmonic 
dispersion law.
However, for a higher carrier density the quantum effects become 
strongly pronounced and result in  a dramatic modification 
of the harmonic spectrum. 
They cause   stiffening of the long-wave length excitations 
and strong softening of the zone boundary magnons [see  Fig. \ref{f3}(b)]. 
%%%%%%%%%%%%%%%%%%%%%%%%%%%%%%%%%%%%%%%%%%%%%%%%%%%%%%%%%%%%%%%%%%%%%
%%%%%%%%%%%%%%%%%%%%%%%%%%%%%%%%%%%%%%%%%%%%%%%%%%%%%%%%%%%%%%%%%%%%%

{\it Ground state phase diagram.} --
As we have already mentioned, there is a critical  carrier
density $n_{cr}$  at which the
 spin wave stiffness vanishes [see Fig. \ref{f2}(a)].
This indicates that there exist another type of magnetic ordering
which competes with FiM order and becomes favorable as $n\rightarrow n_{cr}$.
The most direct route to identify the symmetry of new
magnetic order is to find the wave vector at which spin wave spectrum becomes
unstable. The linear theory can not answer this question, because the  
harmonic dispersion   vanishes  identically in $[111]$  direction, and
no special soft mode is singled out at this order.
The problem can be resolved by  considering quantum effects.
The tendency of the zone boundary
magnons to soften with increasing $n$ [see Fig. \ref{f3}(a) and \ref{f3}(b)]
can  be interpreted as a precursor
effect of a transition to a layered magnetic  ordering with wave-vector
${\bf Q}=\pi/2[1,1,1]$.  The latter 
consists of alternating \{111\} FM planes.
The magnetic moments of neighboring planes can be either misaligned by an angle
$\theta$, which is the case of canted magnetic structure, 
 or aligned antiparallel to  form type-II antiferromagnetic (AFM-II) ordering.
Hence, there are two possible scenarios for the
 phase transition: (a) the transition is
 second order and is to a canted state or 
(b) transition is first order  to
AFM-II ordering. 
To decide which of these two is realized  we have evaluated and compared 
the kinetic energy of
carriers in FM, canted and AFM-II background of core spins.  
We have found that  canted state never minimizes
the energy and hence is never stabilized.

\begin{figure}
\epsfysize=25mm
\centerline{\epsffile{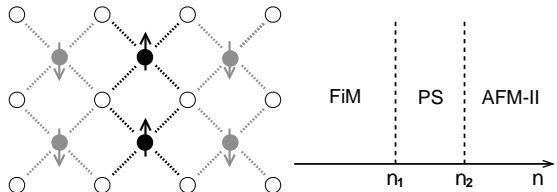}}
\caption{Left panel: Two dimensional (001) cut  of AFM-II structure.
Filled (opened) circles denote Fe (Mo/Re) ions and 
big  arrows stand for  the core spins.
Black (grey) dashed lines indicate the allowed  path for spin down (up) 
electrons; Right panel: ground state phase diagram of model (\ref{H1}).}
\label{f4}
\end{figure}

In Fig. \ref{f4} (left panel) we show
 2D (001) cut of  AFM-II magnetic structure.
In this magnetic structure  carriers  are confined within the
1D FM  stripes and have the dispersion law $\varepsilon_{k}=-2\sqrt{2}t\cos k$
(the factor $\sqrt{2}$ is due to the two possible path of the charge transfer
between nearest-neighbor magnetic ions). 
Figure  \ref{f2}(b) shows the difference between the energies ($\Delta E$)
 of the  AFM-II and FiM  states versus carrier density for 
$\Delta_{\rm CT}=0$ (the same qualitative behavior is realized for finite
 values of $\Delta_{\rm CT}$). As it 
is seen in this figure, with increasing carrier density
$\Delta E$ shows the same
trend, not coincidently, as the spin wave stiffness 
[see Fig. \ref{f2}(a)], 
and vanishes at carrier concentration ${\tilde n}_{cr}$ 
slightly smaller than 
$n_{cr}$ suggested from linear spin wave theory.
For  $n>{\tilde n}_{cr}$  carriers gain more 
kinetic energy in AFM-II state.
This consideration suggests the first order transition from 
ferrimagnetic to AFM-II state. To investigate the possibility of phase
 separation accompanying the first order transition we have evaluated the 
ground state thermodynamic potential $\Omega=E-\mu n$ for homogenous
ferrimagnetic and AFM-II phases as a function of chemical potential $\mu$.
The homogenous state is destabilized with respect to phase separation when 
$\Omega$ of the ferrimagnetic phase becomes larger than that of AFM-II state.
The value of $\mu$ at which this takes place gives the lower $n_{1}$ and upper
$n_{2}$ boundaries of carrier density  for the phase separation regime.
We find that the carrier density $n_1$, at which  which the phase separation 
occurs,  is smaller then the critical carrier densities suggested form the 
above analysis, the  following inequality holds 
$n_1< {\tilde n}_{cr}(n_{cr})<n_2$. For $\Delta_{\rm CT}=0$ we find $n_1\simeq 1.5$ and
$n_2\simeq 2$. 

%%%%%%%%%%%%%%%%%%%%%%%%%%%%%%%%%%%%%%%%%%%%%%%%%%%%%%%%%%%%%%%%%%%%%%%%%%%%%%
%%%%%%%%%%%%%%%%%%%%%%%%%%%%%%%%%%%%%%%%%%%%%%%
Based on  the above  analysis,  in  Fig. \ref{f4} (right panel) 
we present  ground state 
 phase diagram of Hamiltonian (\ref{f1}).
For carrier density $n < n_1$ the ground state is ferrimagnet,
for $n_1<n < n_2$ the systems phase separates (PS) in ferrimagnetic and AFM-II
 phases.
The volume fraction occupied by FiM and AFM-II phases are given by
$v_1=(n-n_1)/(n_2-n_1)$ and $v_2=(n_2-n)/(n_2-n_1)$, respectively.
Here, we point out, that in the real systems, long-range Coulomb interaction,
not considered here, most likely will prevent the phase separation at
 macroscopic scale resulting in the nanoscale domains.\cite{dag}
We have also analyzed the stability of homogenous AFM-II phase within the
 linear spin wave theory and found that this magnetic ordering is indeed
 stable for $n\agt n_2$.  

So far, we have not considered the effect of direct AFM exchange 
$J^{\prime}$  between the
local moments. At the semi-classical level, an effect of $J^{\prime}$ 
is to reduce carrier-induced FM exchange. 
However, in the compounds with large transfer gap $\Delta_{\rm CT}$,  
FM exchange can be largely  suppressed  
and AFM exchange, if strong enough, can stabilize
AFM-II magnetic ordering. Considering 
the metallic compounds with fully polarized carriers
we have also neglected on-site Coulomb repulsion $U$
between charge carriers on $3d$ level of Fe-ions. 
(As at Mo/Re ions electrons reside in more extended $4d/5d$ level
this interaction is much smaller and can be neglected.\cite{sar,sol})
This approximation is fully justified for ${\rm Sr}_{2}{\rm FeMO}_{6}$ 
compound that has one charge carrier per Fe-Mo cell
and hence the Fe-ion is effectively
quarter filled. Therefore the probability of generating doubly-occupied
states is low, and the effect of the Gutzwiller projector on the uncorrelated
wave function is expected to be negligible.
As for the ${\rm Sr}_{2}{\rm FeReO}_{6}$ which has two charge carrier per
formula unit, the on-site $U$, if it is 
not largely screened in the metallic state,
may suppress the charge fluctuations and
renormalize 
electronic caracteristics of the system, 
such as  an effective carrier density and their 
band-width.
However,  as far as magnetic
properties of the systems are concerned, 
we expect  that Coulomb repulsion  can only change  our
quantitative but not qualitative predictions.
The spin wave spectrum as well as its carrier density dependence
 can  be directly verified by inelastic 
neutron scattering experiments on Mo and Re based compounds.

We thank  N. Andrei,  E. I. Kats, D. I. Khomskii, D.E. Logan,
and Ph. Nozi\`eres for their interests and useful discussions. 
Earlier discussion with D. D. Sarma is gratefully acknowledged.  
We are  indebted  to  M. E. Zhitomirsky for continuous valuable discussions
and to T. Ziman for a critical reading of the manuscript. 

%%%%%%%%%%%%%%%%%%%%%%%%%%%%%%%%%%%%%%%%%%%%%%%%%%%%%%%%%%%%%%%%%%%%%

\end{multicols}

\begin{references}
\bibitem[*]{byline} On leave from E. Andronikashvili Institute of Physics,
Georgian Academy of Sciences, Tbilisi, Georgia. 
\bibitem{cmr} 
K.-I. Kobayashi {\it et al}.,  Nature (London) {\bf 395}, 677 (1998);
K.-I. Kobayashi {\it et al}., Phys. Rev. B {\bf 59}, 11159 (1999).
\bibitem{sar} D. D. Sarma {\it et al}., Phys. Rev. Lett. {\bf 85}, 2549 
(2000); 
S. Ray {\it et al}., {\it ibid}.  {\bf 87}, 097204 (2001).
\bibitem{fan} Z. Fang, K. Terakura, and J. Kanamori, 
Phys. Rev. B {\bf 63}, 180407 (2001).
\bibitem{sol}I. V. Solovyev,
Phys. Rev. B {\bf 65}, 144446 (2002).
\bibitem{sai} T. Saitoh {\it et al}.,  Phys. Rev. B {\bf 66}, 035112 (2002).
\bibitem{charge} This is supported by the observation that there are no 
  evidences of  presence of Fe$^{4+}$($d^{4}$) in the
  ground state. 
\bibitem{S} Since completing this work we have become aware of work, 
D. D. Sarma,  Cur. Opin. Solid State Mat. Sci. {\bf 5}, 261 (2001),
  in which a similar point was made.
\bibitem{nagaev} The limit 
$J \rightarrow \infty$ has been   
also discussed  by E. L. Nagaev (cond-mat/0204308), 
however in a different context. 
We thank D. I. Khomskii for pointing this to us.
\bibitem{ferri} H. Kaplan, Phys. Rev. {\bf 86}, 121 (1952).
\bibitem{millis} This is in agreement with  the  dynamical mean-filed
study of a somewhat  extended Hamiltonian by A. Chattopadhyay and 
A. J. Millis, Phys. Rev. B {\bf 64}, 024424 (2001).
\bibitem{gol} To the same order, the analysis for double exchange magnets
has been done in 
D. I. Golosov,  Phys. Rev. Lett. {\bf 84}, 
3974 (2000); N. Shannon and A. V. Chubukov, Phys. Rev. B {\bf 65}, 
104418 (2002)).
\bibitem{dag} E. Dagotto, {\it Nanoscale Phase Separation and Colossal 
Magnetoresistance} (Springer-Verlag, Berlin, 2003).
\end{references}
\end{document}